\documentclass{elsarticle}
\usepackage{graphicx} % Required for inserting images
\usepackage{booktabs}
\usepackage[margin=2.5cm]{geometry}%

\begin{document}

\begin{frontmatter}

\title{Artificial intelligence-driven antimicrobial peptide discovery}
% Advancing Antimicrobial Peptide Discovery through AI-Driven Approaches
% AI-driven Antimicrobial Peptide Discovery
% Artificial intelligence-driven antimicrobial peptide discovery: Challenges and Opportunities 

\author[1]{Paulina Szymczak}
\ead{p.szymczak8@uw.edu.pl}

%\author[1]{Marcin Możejko}
%\ead{marcin.mozejko@uw.edu.pl}

\author[1]{Ewa Szczurek\corref{cor1}}
\ead{szczurek@mimuw.edu.pl}

\cortext[cor1]{Corresponding author}

\affiliation[1]{organization={Faculty of Mathematics, Informatics and Mechanics, University of Warsaw},
addressline={Banacha 2},
postcode={02-097},
city={Warsaw},
country={Poland}}

% Abstract
\begin{abstract}
Antimicrobial peptides (AMPs) emerge as promising agents against antimicrobial resistance, providing an alternative to conventional antibiotics. Artificial intelligence (AI) revolutionized AMP discovery through both discrimination and generation approaches.
The discriminators aid the identification of promising candidates by predicting key peptide properties such as activity and toxicity, while the generators learn the distribution over peptides and enable sampling novel AMP candidates, either \textit{de novo},
or as analogues of a prototype peptide. Moreover, the controlled generation of AMPs with desired properties is achieved by discriminator-guided filtering, positive-only learning, latent space sampling, as well as conditional and optimized generation. 
%In this review, we highlight recent achievements in AI-driven AMP discovery, give an overview of evaluation methods stressing the need for experimenttal validation and standardized benchmarks, and advocate multi-objective optimized generation in AMP design. 
Here we review recent achievements in AI-driven AMP discovery, highlighting the most exciting directions. % give an overview of evaluation methods stressing the need for experimenttal validation and standardized benchmarks, and advocate multi-objective optimized generation in AMP design. 

%By highlighting recent advances, we underscore the importance of AI-driven AMP discovery. 
%This review advocates multi-objective optimization in AMP design and stresses the need for a standardized benchmark. By highlighting recent advances, we underscore the importance of AI-driven AMP discovery. 

%, in analogue mode. 
% including unconstrained and analogue methods,
%such as enhanced activity and reduced toxicity. 
%In this review, we advocate for the incorporation of multi-objective optimization into AMP design and emphasize the importance of establishing a standardized benchmark. This review underscores the significance of AI-driven AMP discovery. 
%Despite challenges like the lack of standardized benchmarking data and effective ranking techniques, recent AI-driven advances in AMP discovery underscore its potential in addressing AMR. 
% This review delves into advancements in AI-driven AMP discovery advancements, highlighting the pressing need for innovative methodologies. 
% This review underscores the significance of AI-driven AMP discovery. 
% The review also underscores the necessity of incorporating multi-objective optimization into AMP design and emphasizes the value of establishing a standardized benchmark. 

\end{abstract}
\end{frontmatter}

% [Introduction]
\section*{Introduction}
% Antimicrobial resistance to existing therapeutics is recognized as a global health and economic hazard~\cite{Murray2022global}. Continuous overuse of antibiotics creates an evolutionary pressure that fuels the emergence and spread of resistant bacterial strains. 
Antimicrobial resistance, fueled by antibiotic overuse that drives the emergence of resistant strains, is recognized as  a global health hazard. It was ranked the third cause of death in 2019, exceeding HIV and malaria~\cite{Murray2022global}.
% By 2050, drug-resistant infections are expected to supersede currently dominating diseases as the main cause of death, with a projected number of 10 million deaths globally per year~\cite{oneill_tackling_2016}. 
% AMR was estimated to be the third leading cause of death in 2019, ahead of both HIV and malaria~\cite{Murray2022global}.
% At the same time, there has been no successful novel class of antibiotics developed since over 30 years~\cite{miethke2021towards}. Therefore, there is a pressing need for the discovery of a new generation of antimicrobial pharmaceuticals.
With no successful novel antibiotics developed for over 30 years~\cite{miethke2021towards}, there is a pressing need for discovering new antimicrobial pharmaceuticals.
% Antimicrobial peptides (AMPs) are intensively investigated as an attractive alternative to conventional antibiotic treatment~\cite{Murray2022global}. 
Antimicrobial peptides (AMPs) are an appealing alternative to known antibiotics~\cite{Murray2022global}.
% As part of natural host defense systems against invading pathogenic microorganisms, AMPs can be active against and kill pathogens that are resistant to antibiotics~\cite{zhu2022antimicrobial}. Crucially, emergence of resistance to AMPs in microbes is slower than to conventional antibiotics~\cite{zhu2022antimicrobial}. 
Innate to host defense systems, they combat antibiotic-resistant pathogens, with slower resistance emergence than conventional antibiotics~\cite{zhu2022antimicrobial}.
% However, in order to be clinically applicable, AMPs not only have to be highly active, but also nontoxic, soluble and stable. 
% While positively charged AMPs demonstrate a pronounced selectivity for negatively charged microbial membranes over neutral eukaryotic cells, many AMPs do display toxicity to mammalian cells. Despite the intensive research on AMP, only 77 AMPs were in clinical trials according to the DRAMP database~\cite{shi_dramp_2022}, but none of them are in clinical use as antibiotics, and only 11 AMPs were commercialized~\cite{carratala_nanostructured_2020}. The main reason why the therapeutic AMP candidates fail clinical trials is that they show comparative to or lower activity than existing antibiotics, high toxicity or low stability.
% This motivates research on innovative computational models for the generation of novel AMPs that would supersede existing AMPs and antibiotics in terms of activity and safety~\cite{magana_value_2020}.
% For clinical use, AMPs must be potent, safe, soluble, and stable.  
Despite extensive research, so far only 77 AMPs were in trials~\cite{shi_dramp_2022}, none serve as clinical antibiotics and only 11 were commercialized~\cite{carratala_nanostructured_2020}. Clinical failures result from low activity, high toxicity, or instability, motivating efforts in designing safer, more effective AMPs~\cite{magana_value_2020}.

% Recent years witnessed an intensive development of artificial intelligence (AI) methods, in particular deep generative models~\cite{bond-taylor_deep_2021}. These models are increasingly applied for drug discovery~\cite{grisoni_chemical_2023} and protein design~\cite{strokach_deep_2022}. 
Recent years witnessed a tremendous advancement in AI, in particular the development of generative and large language models,  revolutionizing the design of drugs~\cite{grisoni_chemical_2023}, proteins~\cite{strokach_deep_2022} as well as AMPs~\cite{melo2021accelerating, chen2022synthetic, fernandes2023geometric, wan2022deep}.
%[ a także 
%https://www.ncbi.nlm.nih.gov/pmc/articles/PMC9598685/ --- MDPI Antibiotics, zarówno traditional i DL
%https://doi.org/10.1016/j.bsheal.2020.08.003 -- 2021, Biosafety and health, Elsevier
%https://doi.org/10.3390/antibiotics10111376 ---Artificial Intelligence and Antibiotic Discovery, MDPI, 2021 można ominąć
% Advances in Antimicrobial Peptide Discovery via Machine Learning and Delivery via Nanotechnology -- MDPI
Since the most recent reviews on AI-driven AMP discovery either %provide more detailed introduction to 
cover the principles of the specific AI methods, including language and generative models~\cite{wan2022deep}, 
%or focus on the potential 
or focus on geometric deep learning methods~\cite{fernandes2023geometric},
here we complement the review of approaches spanning the last two years, highlighting the most exciting directions. 

\begin{figure}[h!]
    \centering
\includegraphics{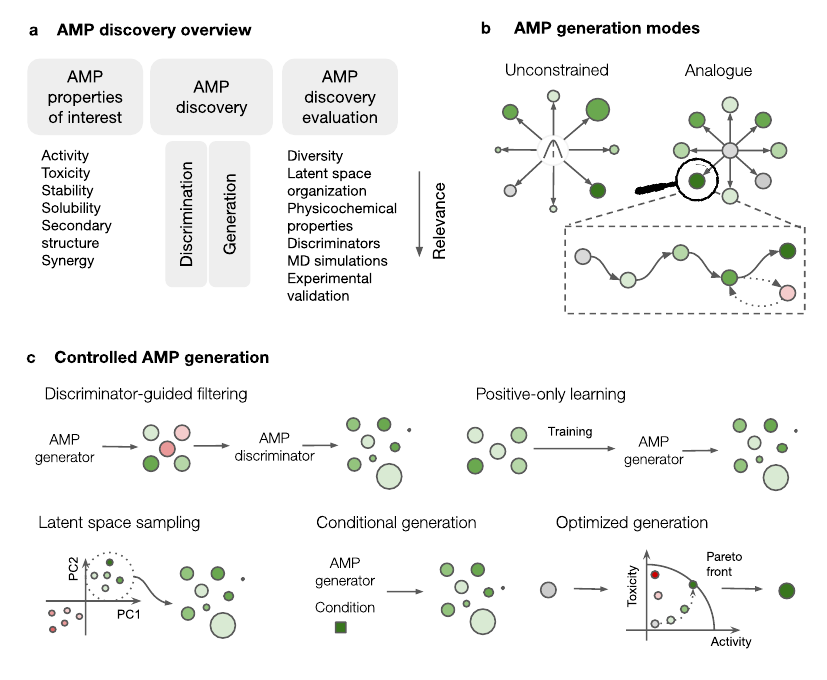}
    \caption{\textbf{AI-driven AMP discovery.} \textbf{a} AMP discovery overview, including AMP properties of interest, main AI methods, and evaluation. The arrow indicates the relevance of evaluation approaches. \textbf{b} AMP generation modes: unconstrained and analogue. In the analogue mode, single step or more steps can be performed, with the latter yielding less similar, but potentially more optimized peptides. \textbf{c} Controlled AMP generation, including: discriminator-guided filtering, positive-only learning, latent space sampling, conditional and optimized generation. Circles represent peptides. The shades of green and red indicate activity and inactivity, respectively. The size of circles are indicative of diverse structural properties e.g. length, secondary structure.}
    \label{fig:fig1}
\end{figure}

We provide a detailed characterization of tasks that the AI methods can perform in AMP discovery,  %outlining grand challenges that still lie ahead. 
introducing the diverse properties of AMPs, and their model representations. We discuss two main categories of AI methods with crucial importance for AMP design: {\textit{discrimination}} and {\textit{generation}} (Figure~\ref{fig:fig1}a). 
% The discriminators predict AMP properties for a given peptide. In turn, generation methods suggest new sequences with desired properties and are often assisted by discriminators. 
We group the most recent discriminators by the predicted properties and categorize the emerging generators by the mode of {\textit{unconstrained}} and \textit{analogue} generation (Figure~\ref{fig:fig1}b). We further discuss approaches to controlled generation of peptides with desired properties (Figure~\ref{fig:fig1}c).
 Moreover, we summarise approaches to evaluation of AMP discovery, both from the methodological and experimental side. Finally, we outline unaddressed challenges impeding AMP delivery to the clinic, highlighting the methodological opportunities for advancement. % and {and 
% Despite the initial successes, multiple methodological obstacles are still in the way~\cite{bender_artificial_2021,bender_artificial_2021-1,van_tilborg_exposing_2022}, and the field needs major breakthroughs. In addition to the generally known complexity of molecular discovery, the task of AMP design has its specific challenges, which are different from the ones typical for small molecule or protein design. These differences result from the physicochemical properties of AMPs,  the type and the availability of data in which these properties are represented, and their heterogeneous modes of action. 
%Similarly to the general field of molecule discovery~\cite{bender_artificial_2021,bender_artificial_2021-1,van_tilborg_exposing_2022},  many challenges still persist in AI-driven AMP design, due to hurdles rooted in AMP properties, data availability, and diverse mode of actions.
% We follow by outlining the real challenges ahead which have so far not been addressed and which still stand in the way of delivering better, potent AMPs to the clinic. 
%Substantial attention is paid to the evaluation of AMP discovery methods both from the methodological and experimental side.

\section*{State of the art}
\subsection*{AMPs are characterized by various properties }

% Substantial efforts in cataloguing AMPs resulted in numerous databases recording sequence (string of amino acids), antimicrobial activity, toxicity, and structure of of antimicrobial peptides, with three most populated and most popular ones (see~\cite{aronica2021computational} for the full list): Database of Antimicrobial Activity and Structure of Peptides (DBAASP)~\cite{pirtskhalava2021dbaasp}, Antimicrobial Peptide Database (APD)~\cite{wang2016apd3}, and Collection of Anti-Microbial Peptides (CAMP)~\cite{gawde2023campr4}. 
AMPs are short (10--100 amino acids) peptides that possess an overall positive charge (typically +2 to +9) and a large percentage ($\geq 30$\%) of hydrophobic amino acids. Although positively charged AMPs preferentially target microbial membranes, some are toxic to mammalian cells.
AMPs are characterized by various properties, each informative of its clinical potential (Figure~\ref{fig:fig1}a), including activity, toxicity, stability, and synergy. 
%[Activity] 
% AMP activity is measured in an antimicrobial assay against a given bacterial strain and predominantly stored as Minimum Inhibitory Concentration (MIC) values, and, less frequently, minimal bacterial concentration (MBC) values. Although it can be assumed that all assays follow the MIC/MBC measurement protocol as described by Clinical and Laboratory Standards Institute (CLSI)~\cite{Wayne2012}, the bacterial strains used in the assays largely vary. Most often MIC is measured against model strains such as \textit{E. coli} ATCC 25922 and \textit{S. aureus} ATCC 25923. Fewer assays include drug-resistant strains, which bring more relevant information for tackling the antimicrobial resistance problem. 
First, activity is measured in antimicrobial assays against various bacterial strains as Minimum Inhibitory Concentration (MIC) or Minimal Bacterial Concentration (MBC). Common reference strains include \textit{Escherichia coli} ATCC 25922 and \textit{Staphylococcus aureus} ATCC 25923, with fewer assays incorporating drug-resistant strains. Assays generally adhere to the MIC/MBC protocol outlined by Clinical and Laboratory Standards Institute (CLSI)~\cite{Wayne2012}. However, they do vary in bacterial concentration and medium usage, potentially introducing bias for CFU levels above $5 \times 10^{5}$ cells/mL due to the inoculum effect~\cite{loffredo2021inoculum}. % for more clinically relevant antimicrobial resistance insights. 
% Antimicrobial assays vary in the bacterial concentration and the medium used, and may be biased for cell densities above colony forming units (CFU) of $5 \times 10^{5}$ cells/mL due to the inoculum effect~\cite{loffredo2021inoculum}.
% DBAASP collects information on both medium and CFU for each activity entry. On top of that, it reports synergistic activity effects between peptides and other antimicrobials, including other antimicrobial peptides, and conventional antibiotics. Synergistic activity is expressed as fractional inhibition concentration (FIC) index, where FIC values less than or equal to $0.5$ mean that the antimicrobial agents have a synergistic effect on each other.
% The activity measurements are reported using either weight/volume ($\mu$g/mL) or molarity (mol/L) unit systems. 
% There is no consensus regarding which unit system should be used, as there are no official guidelines provided by neither CLSI nor the European Committee on Antimicrobial Susceptibility Testing (EUCAST). 

Peptide toxicity is typically measured either as hemolytic activity or cytotoxicity. 
% Hemolytic activity is most often assessed for human erythrocytes and represented as HC50 (the concentration of peptide causing 50\% of hemolysis). However, erythrocytes of other mammals, such as rabbits or sheep, are used very often as well, and hemolytic activity is sometimes reported at other thresholds. Cytotoxicity is tested across various cell types, such as fibroblast, colon, lung, and cancer cell lines. Reported measures include IC50, EC50, and 50\% cell death. 
Hemolysis is mainly evaluated using HC50, indicating the peptide concentration causing 50\% hemolysis in human erythrocytes. Other mammals' erythrocytes, like rabbits or sheep, are also used. Cytotoxicity is examined across diverse cell types (fibroblast, colon, lung, cancer lines) using IC50, EC50, or 50\% cell death.

% Finally, it is possible for one peptide to have multiple different activity or toxicity measurements against the same strain or cell type, taken at different labs. Such measurements can be interpreted as conflicting, depending on the assumed activity/toxicity threshold. In those cases, most common approach is averaging the entries per strain or per genera~\cite{van2021ampgan, losin2021exploring, sharma2023artificial, yan2023deep}. Other strategies include discretization into deciles~\cite{van2021ampgan}, choosing the minimal reported value, discarding peptides with conflicting entries~\cite{teimouri2023bacteria}, or keeping all available measurements despite the conflicts~\cite{ansari2023serverless, ansari2023learning}. 

Numerous databases record activity and toxicity measurements, along with sequence and structure~\cite{aronica2021computational}. 
%Entries of the databases are contributed by individual labs or with the help of large-scale initiatives such as CO-ADD, The Community for Open Antimicrobial Drug Discovery. %~\cite{DOI: 10.1021/acsinfecdis.5b00044.}.
While all databases allow manual browsing, DBAASP~\cite{pirtskhalava2021dbaasp} is the only resource with an Application Programming Interface, facilitating automatic extraction of specific peptide information in desired format. Additionally, DBAASP records medium and CFU data for each activity entry, along with synergistic effects between antimicrobial peptides, conventional antibiotics, and other agents. Synergy is expressed as the Fractional Inhibition Concentration (FIC) index, where FIC values $\leq$ $0.5$ denote significant synergy.
%Conclusion on why working with AMP data is such nightmare based on the previous 

\subsection*{Different AI approaches differ by the way AMPs are represented and provided at their input for training}
% [AMP representation]
% Despite sequence being the default representation method for antimicrobial peptides, artificial intelligence models often use alternative forms of embedding. 
% Indeed, the most prevalent is direct AMP sequence representation, based on single letter amino acid code~\cite{li2022amplify, ghorbani2022deep, ferrell2020generative, das2021accelerated, van2021ampgan, renaud2023latent, ansari2023learning, pandi2022cell, losin2021exploring, buehler2023generative, huang2023identification, capecchi2021machine, surana2021pandoragan, dean2021pepvae, szymczak2023discovering, tucs2023quantum}.
%While sequence is the default for antimicrobial peptides (AMPs), AI models often use alternative embeddings. 
The most prevalent AMP representation based on the amino acid code is simply its sequence~\cite{li2022amplify, ghorbani2022deep, ferrell2020generative, das2021accelerated, van2021ampgan, renaud2023latent, ansari2023learning, pandi2022cell, losin2021exploring, buehler2023generative, huang2023identification, capecchi2021machine, surana2021pandoragan, dean2021pepvae, szymczak2023discovering, tucs2023quantum}.
%Existing methods consider solely standard amino acids, with an exception of MODAN~\cite{murakami2023design} tailored to design peptides with non-proteinogenic amino acids. 
% While N and C-terminal modifications can greatly influence peptide structure and charge, there is only a handful of methods which take that information into account~\cite{sharma2023endl, losin2021exploring}. 
Interestingly, despite the clear influence of N and C-terminal modifications on peptide structure and charge, only a limited number of methods have incorporated this information into their design~\cite{sharma2023endl, losin2021exploring}.
% Recently, sequence embeddings obtained from pretrained language models have been successfully leveraged for AMP discovery~\cite{lee2023ampbert, cao2023designing, sharma2023artificial, salem2022ampdeep, jiang2023explainable, yu2021hmd, ma2022identification, liu2023evolutionary}. 
% On the basis of sequence, other properties are derived and used for modeling, such as amino acid composition or physicochemical properties~\cite{teimouri2023bacteria, vishnepolsky2022comparative, yan2023deep, chharia2021novel, olcay2022prediction, porto2022sense}, sequence similarity~\cite{yang2023ampfinder}, or structural properties, including secondary structure information~\cite{wang2022accelerating}, molecular fingerprints based on SMILES representations~\cite{murakami2023design}, ProtDCal calculated features~\cite{pinacho2021alignment}, as well as atom-type connectivity~\cite{pandey2023samp}.
% Finally, several recent methods methods use a combinations of many encodings~\cite{yang2023ampfinder, sharma2023endl, ansari2023serverless, zhou2023trinet}. 
Beyond sequence, derived AMP properties like amino acid composition, physicochemical attributes~\cite{teimouri2023bacteria, vishnepolsky2022comparative, yan2023deep, chharia2021novel, olcay2022prediction, porto2022sense}, sequence similarity~\cite{yang2023ampfinder}, and structural details like secondary structure~\cite{wang2022accelerating}, molecular fingerprints~\cite{murakami2023design}, and atom-type connectivity~\cite{pandey2023samp} are also used.
Pretrained language model-derived sequence embeddings have recently proven effective for AMP discovery~\cite{lee2023ampbert, cao2023designing, sharma2023artificial, salem2022ampdeep, jiang2023explainable, yu2021hmd, ma2022identification, liu2023evolutionary}. Some methods employ combinations of these encodings~\cite{yang2023ampfinder, sharma2023endl, ansari2023serverless}.

%[Dataset construction]
Most AI-driven AMP discovery methods operate in a supervised setting, with positive and negative datasets used for training.    
However, each study defines these sets differently. 
%[Positive/negative dataset definition
First, the positive examples often constitute the whole collection of peptides from AMP databases, regardless of their target species or strain. This introduces on the one hand large heterogeneity of the positive samples and a bias on the other, as for bacterial AMPs, most of them were tested solely against \textit{E. coli}. 
The negative datasets are frequently sampled from UniProt, excluding entries matching keywords such as antimicrobial, antibiotic, secreted, and similar~\cite{li2022amplify, lee2023ampbert, chharia2021novel, yang2023ampfinder, yu2021hmd, ferrell2020generative, wang2022accelerating, pandi2022cell, ma2022identification}. Yet, Sidorczuk et al.~\cite{sidorczuk2022benchmarks} demonstrated that such negative data selection is highly biased by the sampling methods. 
% Porto et al.~\cite{porto2022sense} proposed an alternative strategy of constructing the negative dataset by shuffling the amino-acid sequences of the positive AMPs and pointed out that deep neural networks perform significantly worse in classifying such data than approaches based on physicochemical descriptors.
Alternatively, Porto {\textit{et al.}}~\cite{porto2022sense} suggested using shuffled AMP sequences as negatives.
%, noting reduced performance of deep neural network-based classifiers compared to physicochemical descriptor-based methods on such designed dataset. 
% A more sophisticated approach to define the positive and negative datasets incorporates the information on activity or toxicity.
% However, there is no consensus what constitutes an active or toxic peptide. Table~\ref{tab:thresholds} summarises the activity and toxicity thresholds assumed by different methods.
 % A more nuanced approach incorporates activity/toxicity data, but consensus on what constitutes active/toxic remains elusive. Activity and toxicity thresholds assumed by different methods are summarized in Table~\ref{tab:thresholds}.
 
A more nuanced approach involves considering activity or toxicity for positive/negative dataset definition, but constructing such datasets is obscured by conflicts in units and contradictory entries in databases. 
First, activity measurements are reported in weight/volume ($\mu$g/mL) or molarity ($\mu$M), and no consensus exists on the preferred unit system due to the absence of guidelines from CLSI.
% When collecting data for training the model, it is a common practice to convert the measurements to uniform units. However, such recalculations usually do not take into account the counterion content in the sample, which was shown to have substantial effect on antistaphylococcal activity and cytotoxicity of antimicrobial peptides~\cite{sikora2018counter, ardino2022impact}. 
Converting measurements to uniform units during data collection is common, but often overlooks the counterion content's influence on antimicrobial peptide activity and cytotoxicity~\cite{ardino2022impact}. Second, one peptide can be reported with varied activity/toxicity measures against the same strain/cell type by different labs. %, potentially causing conflicts in interpretation based on assumed thresholds.
Common approaches of resolving potential conflicts among multiple entries include averaging per strain/genera~\cite{van2021ampgan, losin2021exploring, sharma2023artificial, yan2023deep}, decile discretization~\cite{van2021ampgan}, selecting minimal reported value, discarding conflicting entries~\cite{teimouri2023bacteria}, or retaining all measurements despite conflicts~\cite{ansari2023serverless, ansari2023learning}. Yet, consensus is lacking on what constitutes an active or toxic peptide, with different methods assuming different thresholds (Table~\ref{tab:thresholds}).

\begin{table}[]
\centering
\resizebox{\textwidth}{!}{%
\begin{tabular}{@{}lllll@{}}
\toprule
\multicolumn{1}{c}{\textbf{Ref}} & \multicolumn{1}{c}{\textbf{Task}} & \multicolumn{1}{c}{\textbf{Property}} & \multicolumn{1}{c}{\textbf{Positive}} & \multicolumn{1}{c}{\textbf{Negative}} \\ \midrule
 &  & \textbf{Activity} & \textit{Active} & \textit{Inactive} \\
\cite{vishnepolsky2022comparative} & Discriminaton &  & MIC \textless 25 $\mu$g/ml & MIC \textgreater 100 $\mu$g/ml \\
\cite{teimouri2023bacteria} & Discrimination &  & MIC \textless 25 $\mu$g/ml & MIC \textgreater 100 $\mu$g/ml \\
\cite{das2021accelerated} & Generation &  & MIC \textless 25 $\mu$g/ml & MIC \textgreater 100 $\mu$g/ml \\
\cite{murakami2023design} & Generation &  & MIC \textless 5 $\mu$M & MIC \textgreater 5 $\mu$M \\
\cite{szymczak2023discovering} & Generation &  & MIC \textless 32 $\mu$g/mL & MIC \textgreater 32 $\mu$g/mL \\
\cite{capecchi2021machine} & Generation &  & MIC \textless 32 $\mu$g/mL or 10 $\mu$M & MIC \textgreater 32 $\mu$g/mL or 10 $\mu$M \\
\cite{pandi2022cell} & Generation &  & log MIC \textless 4 $\mu$M & log MIC \textgreater 4 $\mu$M \\
 &  & \textbf{Toxicity} & \textit{Non-toxic} & \textit{Toxic} \\
\cite{capecchi2021machine} & Generation &  & Less than 20\% hemolysis at a concentration of at least 50 $\mu$M & more than 20\% hemolysis at any concentration \\
\cite{ansari2023learning} & Discriminaton &  & HC50 \textgreater 100 $\mu$g/ml & HC50 \textless 100 $\mu$g/ml \\
\cite{ansari2023serverless} & Discriminaton &  & HC50 \textgreater 100 $\mu$g/ml & HC50 \textless 100 $\mu$g/ml \\
\cite{murakami2023design} & Generation &  & HC50 \textgreater 100 $\mu$M & HC50 \textless 100 $\mu$M \\
\cite{sharma2023endl} & Discrimination &  & MHC $\geq$ 50 $\mu$M & MHC  $\leq$ 50 $\mu$M \\
\cite{das2021accelerated} & Generation &  & Hemolytic/cytotoxic activities \textgreater 250 $\mu$g/ml & Hemolytic/cytotoxic activities \textless 200 $\mu$g/ml \\ \bottomrule
\end{tabular}%
}
\caption{\textbf{Activity and toxicity thresholds applied in AMP discovery methods used for defining positive and negative examples.} 
}
\label{tab:thresholds}
\end{table}

% Most frequently peptides with MIC $>$ 100~$\mu$g/mL are classified as inactive and peptides MIC less than or equal to 25~$\mu$g/mL as active~\cite{vishnepolsky2022comparative, das2021accelerated}. A threshold of 32~$\mu$g/mL~\cite{capecchi2021machine, szymczak2023discovering} was also employed. 
% As with antimicrobial activity, there is no clear definition of a toxic peptide. Assumed thresholds for non-toxic peptide include values below 100~$\mu$g/ml~\cite{ansari2023learning, ansari2023serverless}, 200~$\mu$g/ml~\cite{das2021accelerated}, 100~$\mu$M ~\cite{murakami2023design}. There are also methods which take into account both concentration and activity measure~\cite{capecchi2021machine, sharma2023endl}, for example
% peptides reported to cause less than 20\% hemolysis at a concentration of at least 50 $\mu$M as non-hemolytic and the peptides reported to cause more than 20\% hemolysis at any concentration as hemolytic~\cite{capecchi2021machine}.

\subsection*{Artificial intelligence approaches for AMP discovery are dominated by discriminators and generative models}
% Some general introduction here, define major goals: activity, toxicity, solubility, stability
% [TASKS IN AMP]
% EWA TOODO WRITE ON PARRETOO
%The current efforts in AMP discovery are  on finding peptides with high activity, low toxicity, good solubility, and stability. 
Two primary AI-driven tasks for AMP discovery are discrimination and generation (Figure~\ref{fig:fig1}a). %The discriminators predict activity, toxicity, solubility, as well as secondary structure for a given peptide. In turn, generation methods suggest new sequences with desired properties and are often assisted by discriminators. 
We begin by describing models in the discrimination category.

\subsubsection*{AMP discriminators}
% [AMP classification]
% Binary AMP classification
The field is flooded with methods that classify peptides broadly as either AMP or non-AMP. %Some methods for this task are based on hand-engineered descriptors, such as ones describing physicochemical properties~\cite{porto2022sense}. 
Recent approaches of this type, such as AMPlify~\cite{li2022amplify}, a bidirectional long short-term memory (LSTM) model with additional multihead attention mechanism, or AMP-BERT~\cite{lee2023ampbert}, a BERT model pretrained on protein sequences, predominantly use deep learning to automatically derive descriptive features for classification. %For example, 
%Large language models such as are also employed for AMP classification.
%An interesting approach was proposed in  VGG16-AMP~\cite{pandey2023samp}, where information on sequence and structure is converted into 3-channel image based on connectivity matrices encoding atom-type angle, dihedral, and bond connectivity, and methods designed for image recognition were applied for the AMP classification task. %Representing peptide structure as an image allows the usage of methods and models designed specifically for image recognition.
% VGG16-AMP~\cite{pandey2023samp} converted information on sequence and structure into 3-channel image based on connectivity matrices encoding atom-type angle, dihedral, and bond connectivity, thereby making methods designed for image recognition applicable to the AMP classification task.
VGG16-AMP~\cite{pandey2023samp} converts sequence and structure information into a 3-channel image based on connectivity matrices, enabling the application of image recognition methods. 
% While AMP classification is a problem tackled by the largest number of studies, it is also the least specific and least useful for AI-guided AMP discovery, and thus we do not provide a comprehensive survey of these studies.
% A peptide classified as simply an AMP is not guaranteed to have high activity against a specific strain of bacteria.
% %Firstly, a peptide classified as AMP by such a method is not guaranteed to work against a specific strain of bacteria. 
% %Secondly, having a high probability of being an AMP is not equivalent with having high antimicrobial activity. 
% Our evaluation of AMP classification methods~\cite{szymczak2023discovering} showed high false positive rates and low predictive performance for peptides known to be highly active against \textit{E. coli} strains. Interestingly, a method based on hand-engineered descriptors~\cite{porto2022sense} was proven to work competitively to deep learning approaches. 
% AMP multi-label classication 
Antimicrobial peptides are further subcategorized into antibacterial, antiviral, antifungal, anticancer (and more).
% Methods were proposed that classify AMPs into specific subgroups, such as anticancer peptides~\cite{yang2023cacpp}. However, recent methods aim to address a more ambitious challenge of multi-label classification, which is largely obscured by the scarcity of training data~\cite{yu2021hmd, yang2023ampfinder}. 
While some methods classify AMPs into specific subcategories like anticancer peptides in a binary fashion~\cite{yang2023cacpp}, recent approaches tackle the challenging multi-label classification, hindered by limited training data~\cite{yu2021hmd, yang2023ampfinder}. For example, Yang et al.~\cite{yang2023ampfinder}  propose a two-stage binary classification: first into AMP/non-AMP categories, and second, antimicrobial/non-antimicrobial subcategories. Notably, in these basic classification tasks, hand-engineered descriptor-based methods, %like Porto et al.'s, 
offer competitive performance compared to deep learning approaches~\cite{porto2022sense}.
 % In~\cite{yang2023cacpp} contrastive learning is used to improve the classification performance.   
 However, AMP or AMP subcategory classification, addressed by numerous studies, lacks specificity and utility for AI-driven AMP discovery. Peptides classified as AMPs might not exhibit high activity against specific bacterial strains, as indicated by our evaluation~\cite{szymczak2023discovering}.

MIC prediction methods have the potential to give more specific and practically applicable predictions.
% [MIC prediction]
% MIC prediction is stated either as a regression problem, where specific MIC values are returned~\cite{yan2023deep, sharma2023artificial}, or as a classification problem, where active peptides are separated from inactive ones~\cite{teimouri2023bacteria, vishnepolsky2022comparative} based on an arbitrary threshold. 
MIC prediction is framed either as regression, yielding MIC values~\cite{yan2023deep, sharma2023artificial}, or as classification, discerning active from inactive peptides~\cite{teimouri2023bacteria, vishnepolsky2022comparative} using a set threshold. 
% As mentioned previously, it is not straightforward to first define the activity threshold, and second, how to integrate MIC measurements which differ both in terms of the target strain and experimental conditions.  
%Defining the threshold and harmonizing varying MIC measurements across strains and conditions pose challenges.
% Comparing activity with existing antibiotics such as imipenem~\cite{das2021accelerated},ampicillin~\cite{tucs2023quantum}, polymyxin B~\cite{ma2022identification} or colistin~\cite{lata2023evolutionary} also has been used.
Depending on the dataset construction, the MIC prediction methods provide predictions for specific 
%Gram staining type, 
genera~\cite{losin2021exploring, sharma2023artificial}, species~\cite{teimouri2023bacteria}, or strain~\cite{vishnepolsky2022comparative}. 
% Microbial strain specific MIC prediction 
%The activity prediction was also tackled by 
More general, methods for microbial specific strain (MSS) prediction~\cite{vishnepolsky2022comparative, sharma2023artificial} predict activity for a given peptide-bacterial strain pair. These approaches leverage the genomic information of the strain, including inter-strain similarity as well as oligonucleotide composition.
% [AMP comparison]
Losin and Veltri introduced an approach for activity comparison~\cite{losin2021exploring}, devising a model employing Siamese neural networks, which predicts the difference in MIC for each AMP pair. 
% [Synergy prediction]
Finally, Olcay et al.~\cite{olcay2022prediction} proposed a method which predicts synergistic effects as FIC for a peptide and an antibiotic against a given bacterial strain.

Compared to AMP and activity, much fewer discriminators for other properties were developed. 
% [Toxicity prediction]
In particular, classifiers of peptides either as toxic/non-toxic or hemolytic/non-hemolytic were proposed~\cite{salem2022ampdeep, sharma2023endl, ansari2023serverless, ansari2023learning}. %Again, toxicity prediction can be treated both as a classification problem or a regression problem. 
% Binary toxicity classification
% HC50 regression 
%Alternatively, specific hemolytic activity values can be predicted, but those method are extremely scarce. 
In a recent study, Salem et al.~\cite{salem2022ampdeep} used transfer learning, first teaching a large language model to recognize secretory peptides, and then applying such model to hemolytic activity classification. 
% [Solubility prediction]
AI is used as well to predict the solubility of peptides~\cite{ansari2023serverless, wu2021epsol, chen2021structure}. However, those methods are applied to longer sequences and may not be as accurate for shorter peptides. %Ansari and White~\cite{ansari2023serverless} proposed an approach for estimating relative ease of solid-state peptide synthesis using Milton Coupling Efficiency. 
Finally, a recent secondary structure classifier tailored for short peptide sequences was built upon a combination of a pretrained language model, hypergraph multihead attention network, and bi-LSTM with conditional random fields (CRF)~\cite{jiang2023explainable}. Apart from structure classification, structure prediction was also applied to AMPs~\cite{jumper2021alphafold, pandi2022cell}, although our evaluation indicated that it tends to overly stable secondary structures~\cite{szymczak2023discovering}. Similarly as in case of solubility, most of the available models for structure prediction focus on large proteins and might not be suitable for AMPs. 
% [Secondary structure prediction]
%Apart from solubility, secondary structure is often considered as an important proxy for membrane interaction and as input for molecular dynamics (MD) simulations. While AlphaFold~\cite{jumper2021alphafold} has been used for obtaining peptide structures~\cite{pandi2022cell}, we have shown that it tends to overly stable secondary structures~\cite{szymczak2023discovering}. Similarly as in case of solubility, most of the available models for structure prediction focus on large proteins and might not be suitable for AMPs. 

%fractional inhibitory concentration  as synergism no interaction

%EWA TODO --------
\subsubsection*{AMP generators}

% Data mining 
One traditional way of discovering AMPs was to scan a given database of peptides using discriminators. Similar approaches are still explored today, for example via an exhaustive screen of large peptide libraries~\cite{huang2023identification}, or by mining peptides from metagenomic data~\cite{ma2022identification}. However, most recent studies aim to leverage generative AI for AMP discovery (Table~\ref{tab:generation}). 
% AMP unconstrained i analogue generation
%We group AMP generation methods with respect to the mode of generation into {\textit{unconstrained}} and \textit{analogue} generation % and {\textit{optimized generation}} 
%(Fig.~\ref{fig:fig1}b). 
In unconstrained generation, peptides are freely sampled from the model {\textit{de novo}}~\cite{szymczak2023discovering, surana2021pandoragan, capecchi2021machine, buehler2023generative, ghorbani2022deep, pandi2022cell, van2021ampgan, wang2022accelerating, das2021accelerated, ferrell2020generative}. 
In contrast, analogue generation takes a given peptide as a prototype and generates its analogues~\cite{szymczak2023discovering, dean2021pepvae, renaud2023latent, murakami2023design, tucs2023quantum, hoffman2022optimizing, liu2023evolutionary}. Analogue generation can proceed in a single or several steps, with more steps yielding peptides that are potentially more optimized but less similar to the prototype.  
% AMP de novo i optimized generation
%Finally, in optimized generation mode, the model idealises  a peptide with respect to some assumed property(ties) in an iterative manner
%~\cite{murakami2023design, tucs2023quantum, hoffman2022optimizing, liu2023evolutionary}.
%A schematic representation of the generation modes are captured on Fig.~\ref{fig:fig1}b. 

From the modeling framework perspective, the most popular approaches to AMP generation are GANs~\cite{ferrell2020generative, van2021ampgan, cao2023designing, surana2021pandoragan} and VAEs~\cite{wang2022accelerating, pandi2022cell, ghorbani2022deep, renaud2023latent, hoffman2022optimizing, dean2021pepvae, tucs2023quantum, szymczak2023discovering}. The unconstrained generation mode is implemented by all GAN-based models, but also
%, in particular their bidirectional conditional variant~\cite{van2021ampgan, ferrell2020generative, pando}%. 
%The unconstrained generation problem was tackled as well using
Recurrent Neural Network (RNN)~\cite{capecchi2021machine}, and Graph Neural Network (GNN)~\cite{buehler2023generative}, as well as all VAE models~\cite{das2021accelerated, wang2022accelerating, pandi2022cell, szymczak2023discovering}. 
%Similarly to conditional GANs, 
%conditional VAEs (cVAEs) was used as well for unconstrained AMP generation in HydrAMP~\cite{szymczak2023discovering}. 
The analogue generation task is performed solely using VAEs~\cite{dean2021pepvae, renaud2023latent, szymczak2023discovering, hoffman2022optimizing, tucs2023quantum, jain2022biological}, where it is possible to encode the prototype peptide and obtain its analogues through sampling process in the latent space.
%and output from the decoder

%[The need of controlled generation]
%All the generative models in general aim to generate novel antimicrobial peptides. However, in order to obtain peptides with desired AMP properties, %high activity, low toxicity, and desired stability or structure, 
In order to obtain peptides with desired AMP properties it is essential to control the generation process with respect to these properties (Figure~\ref{fig:fig1}c). The controlled properties of peptides include AMP~\cite{das2021accelerated, cao2023designing, szymczak2023discovering, renaud2023latent, jain2022biological}, activity~\cite{surana2021pandoragan, ferrell2020generative, das2021accelerated, van2021ampgan, pandi2022cell, capecchi2021machine, szymczak2023discovering, dean2020variational, hoffman2022optimizing, tucs2023quantum},  toxicity~\cite{das2021accelerated, capecchi2021machine, hoffman2022optimizing, tucs2023quantum}, microbial target~\cite{ferrell2020generative, van2021ampgan}, target mechanism~\cite{ferrell2020generative, van2021ampgan}, hydrophobicity~\cite{renaud2023latent}, secondary structure~\cite{wang2022accelerating, buehler2023generative, das2021accelerated},  as well as sequence length~\cite{ferrell2020generative, van2021ampgan}. 
The controlled AMP generation is obtained by approaches including \textit{discriminator-guided filtering}, \textit{positive-only learning}, \textit{latent space sampling}, \textit{conditional generation} and \textit{optimized generation} (Figure~\ref{fig:fig1}c). 

First, in discriminator-guided filtering, the generative models are coupled with discriminators~\cite{das2021accelerated, pandi2022cell, cao2023designing, capecchi2021machine} to select peptides with desired properties. Most often such discriminators are trained on the independent training set and applied on the generated samples. In CLaSS~\cite{das2021accelerated}, the discriminators were trained in the latent space of the generative model in a rejection sampling scheme. %Discriminators were also used to define and evaluate conditions, additional variables encoding properties of peptides in the latent space of our recent generative model called HydrAMP~\cite{szymczak2023discovering}.
%Another usage of classifiers is conditional generation~\cite{szymczak2023discovering}, where classifiers guide the generation process to sample peptides with desired properties~\cite{szymczak2023discovering}. 

%Apart from discriminator-guided filtering, g
%Generation of active peptides is achieved as well by %deliberate construction of the training dataset in 
%positive-only learning, where some approaches, predominantly Generative Adversarial Networks (GANs)~\cite{surana2021pandoragan, cao2023designing}, train exclusively on positive examples. S
Positive-only learning is predominantly implemented by Generative Adversarial Networks (GANs)~\cite{surana2021pandoragan, cao2023designing} and corresponds to training exclusively on positive examples.
Since these models learn the underlying data distribution, we can expect that the generated peptides will also be positive. %operate in a semi-supervised setting, with positive examples only. 
%In PandoraGAN~\cite{surana2021pandoragan} generation of novel antiviral peptides is achieved via positive-only learning on experimentally validated peptides with high antiviral activity. 
For example, PandoraGAN was trained on experimentally validated peptides with high antiviral activity~\cite{surana2021pandoragan}. 
Positive-only learning was also combined with transfer learning, with recurrent neural network (RNN) model first trained on active peptides and later fine-tuned using active and non-hemolytic peptides~\cite{capecchi2021machine}. 

%Positive unlabeled learning was also employed, where the model was presented with a small set of positive samples and a much larger set of unlabeled samples, containing both positive and negative samples~\cite{ansari2023learning}. 

Latent space sampling is performed by generative models equipped with the latent space, leveraging the structure of this space to sample candidates with desired properties~\cite{wang2022accelerating, dean2021pepvae, renaud2023latent}. For example, in PepVAE~\cite{dean2021pepvae} active peptides were sampled from latent space regions which are most distant in terms of cosine similarity to an inactive query peptide. In turn, Renaud et al.~\cite{renaud2023latent} introduced PCA property aligned sampling, where by fixing a principal component of interest (e.g. identified as correlated with hydrophobicity) it is possible to generate similar peptides by sampling nearby points along other principal components. 
Finally, Wang et al.~\cite{wang2022accelerating} encoded information on both the sequence and structure in the same discrete latent space of multi-scale vector quantized-VAE, which allowed generation of peptides with desired structure. 

Conditional generation models are equipped with additional variables encoding for conditions corresponding to selected AMP properties, and trained to generate samples satisfying these conditions. Conditional generators include GAN-based~\cite{ferrell2020generative, van2021ampgan} and GNN~\cite{buehler2023generative} models. We have recently proposed HydrAMP~\cite{szymczak2023discovering}, an extended conditional VAE, coupled with a pair of classifiers.
%, which allowed generation of active analogues of a query peptide. 
Because HydrAMP was trained to perform analogue generation in temperature-controlled setting, it was possible to generate highly active analogues of a peptide without any antimicrobial activity. Morevoer, HydrAMP is the only model capable of both unconstrained and analogue generation.

Finally, optimized generation models aim to modify a given query peptide towards improved properties~\cite{ tucs2023quantum, jain2022biological}. State-of-the-art optimization algorithms were applied for AMP design, such as % Gaussian process-based surrogate models in 
Bayesian optimization~\cite{murakami2023design} or multi-objective  evolutionary algorithm~\cite{liu2023evolutionary}. %combined with decomposition-based framework with an elite archive and local search~\cite{liu2023evolutionary}. %Gaussian process-based surrogate models were used in Bayesian optimization of a score computed as 
%total probability of improvement (TPI) which is a product of PI of six antimicrobial and hemolytic activities. 
%a product of a probability of improvement of  antimicrobial activities against six strains of bacteria and hemolicity~\cite{murakami2023design}.
%In turn, Liu etal.~\cite{liu2023evolutionary} optimized peptides using an evolutionary algorithm in a decomposition-based framework with an elite archive and local search. 
However, emerging AI-driven approaches combine generative models with optimization steps. For example, Hoffman et al.~\cite{hoffman2022optimizing} introduced a VAE-based model with gradient descent zeroth-order optimization to convert a toxic peptide into a non-toxic one, while maintaining antimicrobial properties. In binary VAE~\cite{tucs2023quantum} each peptide was scored by a distance from the Pareto front via non-dominated sorting, which was followed by optimization of the prediction score via quantum annealing. As a proof of concept, the model was used to optimize easily calculable properties, such as charge, density, instability index, and Boman index. 
Finally, Jain et al.~\cite{jain2022biological} proposed an active learning algorithm leveraging epistemic uncertainty estimation and GFlowNets as a generator of a diverse batch of candidates for active peptides.

\begin{table}[h!]
\centering
\resizebox{\textwidth}{!}{%
\begin{tabular}{lllllllll}
\hline
\textbf{Reference} & \textbf{Generation mode} & \textbf{\begin{tabular}[c]{@{}l@{}}Generation \\ framework\end{tabular}} & \textbf{\begin{tabular}[c]{@{}l@{}}Controlled \\ condition\end{tabular}} & \textbf{Discriminators} & \textbf{\begin{tabular}[c]{@{}l@{}}Approach to \\ controlled generation\end{tabular}} & \textbf{\begin{tabular}[c]{@{}l@{}}Experimental\\ validation\end{tabular}} & \textbf{MD} & \textbf{Details} \\ \hline
\cite{surana2021pandoragan} & Unconstrained & GAN & Antiviral activity &  & Positive-only learning & no & no &  \\
\cite{cao2023designing} & Unconstrained & GAN & AMP & yes & \begin{tabular}[c]{@{}l@{}}Discriminator-guided\\ filtering\end{tabular} & yes & yes &  \\
\cite{van2021ampgan} & Unconstrained & \begin{tabular}[c]{@{}l@{}}Bidirectional \\ cGAN\end{tabular} & \begin{tabular}[c]{@{}l@{}}Sequence length, \\ microbial target, \\ target mechanism, \\ activity\end{tabular} &  & Conditional generation & no & no &  \\
\cite{ferrell2020generative} & Unconstrained & \begin{tabular}[c]{@{}l@{}}Bidirectional \\ Wasserstein cGAN \\ with gradient \\ penalty\end{tabular} & \begin{tabular}[c]{@{}l@{}}Sequence length, \\ microbial target, \\ target mechanism, \\ activity\end{tabular} &  & Conditional generation & yes & yes &  \\
\cite{capecchi2021machine} & Unconstrained & RNN & Activity, toxicity & yes & \begin{tabular}[c]{@{}l@{}}Positive-only learning, \\ Discriminator-guided\\ filtering\end{tabular} & yes & no &  \\
\cite{buehler2023generative} & Unconstrained & \begin{tabular}[c]{@{}l@{}}Multitask \\ autoregressive\\ transformer \\ GNN\end{tabular} & Secondary structure &  & Conditional generation & no & no & \begin{tabular}[c]{@{}l@{}}Forward and inverse\\ training\end{tabular} \\
\textbf{\cite{pandi2022cell}} & \textbf{Unconstrained} & \textbf{VAE} & \textbf{Activity} & \textbf{yes} & \textbf{\begin{tabular}[c]{@{}l@{}}Discriminator-guided\\ filtering\end{tabular}} & \textbf{yes} & \textbf{yes} & \textbf{Cell-free biosynthesis} \\
\cite{das2021accelerated} & Unconstrained & \begin{tabular}[c]{@{}l@{}}Wasserstein \\ autoencoder\end{tabular} & \begin{tabular}[c]{@{}l@{}}AMP, activity, \\ toxicity, structure\end{tabular} & yes & \begin{tabular}[c]{@{}l@{}}Discriminator-guided\\ filtering\end{tabular} & yes & yes & \begin{tabular}[c]{@{}l@{}}Classifiers trained\\ in the latent space\end{tabular} \\
\cite{wang2022accelerating} & Unconstrained & \begin{tabular}[c]{@{}l@{}}Vector \\ quantized VAE\end{tabular} & Secondary structure &  & Latent space sampling & yes & no & Discrete latent space \\
\textbf{\cite{szymczak2023discovering}} & \textbf{\begin{tabular}[c]{@{}l@{}}Unconstrained,\\ analogue\end{tabular}} & \textbf{cVAE} & \textbf{AMP, activity} & \textbf{yes} & \textbf{Conditional generation} & \textbf{yes} & \textbf{yes} & \textbf{\begin{tabular}[c]{@{}l@{}}Temperature \\ controlled creativity\end{tabular}} \\
\cite{dean2021pepvae} & Analogue & VAE & Activity &  & Latent space sampling & yes & no & \begin{tabular}[c]{@{}l@{}}Sampling based\\ on cosine similarity\end{tabular} \\
\textbf{\cite{renaud2023latent}} & \textbf{Analogue} & \textbf{\begin{tabular}[c]{@{}l@{}}VAE-like models \\ (RNN, RNN \\ with attention, \\ Wasserstein \\ autoencoder, \\ adversarial \\ autoencoder, \\ transformer)\end{tabular}} & \textbf{\begin{tabular}[c]{@{}l@{}}AMP, \\ hydrophobicity\end{tabular}} & \textbf{} & \textbf{Latent space sampling} & \textbf{no} & \textbf{no} & \textbf{\begin{tabular}[c]{@{}l@{}}PCA property\\ aligned sampling\end{tabular}} \\
\textbf{\cite{hoffman2022optimizing}} & \textbf{Analogue} & \textbf{VAE} & \textbf{Activity, toxicity} & \textbf{yes} & \textbf{Optimized generation} & \textbf{no} & \textbf{no} & \textbf{\begin{tabular}[c]{@{}l@{}}Zeroth-order\\ optimization, \\ gradient descent\end{tabular}} \\
\textbf{\cite{tucs2023quantum}} & \textbf{Analogue} & \textbf{Binary VAE} & \textbf{Activity, toxicity} & \textbf{yes} & \textbf{Optimized generation} & \textbf{yes} & \textbf{no} & \textbf{\begin{tabular}[c]{@{}l@{}}D-Wave quantum \\ annealer, non-dominated\\ sorting, factorization \\ machine\end{tabular}} \\
\cite{jain2022biological} & Analogue & GFlowNets & AMP & yes & Optimized generation & no & no & \begin{tabular}[c]{@{}l@{}}Active learning, epistemic \\ uncertainty\end{tabular} \\ \hline
\end{tabular}%
}
\caption{\textbf{An overview of generation methods applied in AMP discovery}. The table summarises the generation mode, modeling framework, the controlled conditions, the usage of discriminators, the approach to controlled generation, as well as whether experimental validation, and MD simulations in the presence of membrane were carried out. The rows marked in bold indicate methods of special and outstanding interest.}
\label{tab:generation}
\end{table}

\subsubsection*{Evaluation}
% For proper assessment of utility of a given AI method for a given AMP design task, informative evaluation measures are essential. In case of the discriminative models, most of the standard evaluation measures are applied, such as 
% %precision, recall, FDR, or AUROC for the classifiers, and RMSE or correlation for the regressors. 
% area under the curve (AUROC) for the classifiers or root mean squared error (RMSE) for the regressors. 
For a thorough assessment of the utility of an AI method in AMP design tasks, informative evaluation measures are vital.
In the case of discriminative models, standard evaluation metrics such as area under the curve (AUROC) for classifiers or root mean squared error (RMSE) for regressors are commonly used. 
However, there is no SOTA method that each new method would be compared to. 
%A standard benchmarking dataset is also missing. 
Moreover, due to the fact that every discriminative model is trained on a different dataset regarding activity thresholds, similarity cut-offs, construction of the negative dataset etc., each model should be re-trained on the same dataset for the sake of comparison. 

%Boring: general generation property such as diversity
Compared to the discriminators, the evaluation of generation methods is generally much trickier~\cite{bilodeau2022generative}. 
In case of AMPs, the additional difficulty stems from the fact that every method is trained on a different dataset and targets different peptide features and that these features are hard to estimate or predict computationally. 
%Moreover, in case of generative method there is a trade-off between reconstruction and creativity. Additionally, there is no standard metric which could be evaluated in the benchmarking. This approach is broadly used in small molecule design, where the models are trying to obtain a sample with a better metric (usually logP, a metric of…) than the current optimum in the test set. 
%Currently, most of the generative methods are evaluated internally and/or externally. 
%The external validation focuses on comparison with other generative models. 
% One standard property that is evaluated for any generative model is  \textit{diversity}, which can be understood as the ability to produce multiple samples that show some differences from each other and from the training data. 

Diversity is a basic criterion for assessing generative models, reflecting their ability to produce distinct samples from both each other and training data~\cite{jain2022biological}. 
Common metrics for quantifying diversity in AMP generators are Levehnstein distance~\cite{wang2022accelerating, capecchi2021machine, szymczak2023discovering}, BLEU~\cite{ghorbani2022deep}, Jaccard similarity~\cite{renaud2023latent}, as well as pairwise sequence similarity score~\cite{das2021accelerated, hoffman2022optimizing, renaud2023latent, pandi2022cell}. 
%However, Liu et al.~\cite{liu2023evolutionary} criticize both Levehnstein and Jaccard metrics for ignoring the order of amino acids and therefore they should not be used to evaluate similarity of AMPs. They propose their own metric based on the number of common amino acids between two peptides as well as the total length of common subsequences between two peptides.
However, Liu et al.~\cite{liu2023evolutionary} criticize both Levehnstein and Jaccard metrics for ignoring the order of amino acids and propose their own metric based on common amino acids and shared subsequences between peptides.

%Specific case of latent space models
The specific subclass of latent space-based generative models is evaluated with respect to reconstruction ability, interpretability, as well as organization of the latent space~\cite{renaud2023latent}.

%Also boring: discriminators or psychicochemical props
% For the specific task of AMP design, it is less crucial to generate a large number of diverse samples than to produce (potentially less) AMPs with desired features, such as a high probability of being AMP, high activity, and low toxicity. 
For AMP design, generating samples with specific desired properties (e.g., high AMP probability, activity, and low toxicity) is more important than their diversity. 
With this respect, generative models are evaluated by applying discriminators on the generated peptide sequences and using their predictions as proxies of the true properties~\cite{das2021accelerated, pandi2022cell, cao2023designing, capecchi2021machine}. Here, the limited accuracy of the existing discriminators may bias or provide overly optimistic evaluation. 
% Physicochemical properties of obtained peptide sequences, such as amphipathicity, hydrophobicity, or charge, are much easier to compute for any given peptide, but are also much less specific indicators of the desired features. 
In contrast to AMP properties, physicochemical properties like amphipathicity, hydrophobicity, and charge are easier to compute but less specific indicators of desired features for peptide sequences. 

Molecular dynamics (MD) simulations have the potential to deliver more reliable evaluation measures, at the cost of high simulation time and were used  for evaluation of some of the generators (Table~\ref{tab:generation}). MD was used to study both the stability of the secondary structure~\cite{cao2023designing}, and the mode of action of peptides~\cite{capecchi2021machine}. Recently, MD simulations emerged as a proxy of interaction with bacterial membrane, with scores assigned to each sequence~\cite{ferrell2020generative, szymczak2023discovering}. 
% For example, Ferrell et al.~\cite{ferrell2020generative} proposed a score correlated to membrane-binding propensity based on free energy change ($\Delta$G). 
% %It compares a candidate sequence to a known membrane-active peptide, magainin 2 ($\Delta$G+) and a known non-AMP sequence ($\Delta$G-). 
% We have introduced an MD-based descriptor \textit{S}, describing the level of peptide burial within the membrane~\cite{szymczak2023discovering}. It reflects the fraction of peptide heavy atoms buried below the membrane surface and is indicative of the peptides’ ability to bind and penetrate the membrane core. 
MD was also used to investigate AMP preference of bacterial over human membrane~\cite{pandi2022cell}.
%contrast peptides with respect to their behaviour in the presence of bacterial and human membrane~\cite{pandi2022cell}. 
%investigated AMP preference of bacterial over human membrane. distance of peptide center of mass to membrane midplane. 
These approaches were all based on computationally-heavy fully atomistic simulations. An alternative might be coarse grained simulations, which are much faster, at the price of  specificity drop~\cite{das2021accelerated}.
% These approaches were all based on computationally-heavy fully atomistic simulations. An alternative might be coarse grained simulations, which are much faster. This approach was used by Das et al.~\cite{das2021accelerated}, where the variance in the number of contacts between positive residues and membrane lipids is treated as a discriminator of antimicrobial activity. This measure, however, showed a very low specificity of only 63\%. 

%Eyes flushihng: experimental validations!
The ultimate evaluation of generated peptides should be performed in an experimental setting, where activity (MIC), toxicity (HC50) and other peptide features are measured on target bacteria or cell lines. Not all published generators were validated experimentally (Table~\ref{tab:generation}). Since there is no consensus on MIC thresholds (see Table~\ref{tab:thresholds}), we proposed to summarize experimental activity validation using an AMP success rate curve instead of a single value~\cite{szymczak2023discovering}. The AMP success rate curve displays the fraction of active peptides among all experimentally tested ones, as a function of the activity threshold. 
% It is important to note, that the most indicative of the clinical potential of the generated peptides is their success rate evaluated at very low AMP thresholds, such as MIC of 
% %32~$\mu$g/mL and lower.
% 4~$\mu$g/mL and lower, making their activity more comparable to antibiotics on the market. 
Crucially, the clinical potential of peptides is best assessed at low thresholds, e.g., MIC $\leq$ 2~$\mu$g/mL or 1~$\mu$M, akin to market antibiotics.
Only few generative models report such high activity of their generated peptides~\cite{szymczak2023discovering, pandi2022cell, dean2021pepvae}.
% However, with only very few exceptions~\cite{szymczak2023discovering, pandi2022cell, dean2021pepvae, capecchi2021machine, tucs2023quantum}, generative models do not report such high activity of their generated AMPs.
%an appropriate measure of %In case of AMP design, such metrics could be activity understood as MIC and/or toxicity (HC50), however, it is not possible to obtain the score without wet-lab validation. 
%It is straightforward that the ultimate goal of AMP discovery is to generate a peptide which could have clinical application. In fact, predictions in conservative range below 32~$\mu$g/mL should be evaluated. Therefore, the crucial metric reported should be the AMP discovery rate, as in ~\cite{szymczak2023discovering} which measures the fraction of peptides verifiers as AMP among all proposed candidates. Again, this is only possible once wet-lab validation is included in the experiment design. 
%Finally, the wet-lab experiments include predominantly MIC and MBC measurements against at least one microbial strain, and hemolitic assays in which HC50 is reported. 
Once promising peptides are identified in initial screening of their activity and toxicity, more in depth experimental characterization should follow. This should include antimicrobial assays against drug resistant strains,
%such as \textit{A. baummanni}, \textit{P. aeruginosa}, or carbapenem-resistant \textit{K. pneumoniae} (carrying carbapenemase genes such as KPC, NDM-1 and OXA-48) 
%as well as other microbes such as fungi. 
%(e.g. Candida albicans ATCC 90028 and Candida neoformans ATCC 208821). 
% Additionally, not only hemolytic activity (HC50) but also cytotoxicity against HEK293 and other cell lines should be investigated, along with the investigation of the ability of membrane disruption, time killing as well as resistance induction assays to measure the time scale
% within which the peptide remains active and the time needed for the bacteria to acquire resistance mechanisms. 
 cytotoxicity on cell lines, membrane disruption, time killing, resistance induction, biofilm efficacy and others. 
% Finally, additional possible experiments include  biofilm efficacy assays to test whether the discovered AMPs may modify the biofilm
% production, a phenomenon, which may cause chronic, nosocomial, and medical device-related infections in
% the clinic [refs].
% Due to the fact that the laborious wet-lab validation, in particular the solid-state synthesis, is the limiting factor in fast identification of novel peptides, faster experimental approaches were proposed. For example, Pandi et al.~\cite{pandi2022cell} combined the VAE-based generation with fast cell-free synthesis.

%Latent-space based models can be evaluated with respect to reconstruction ability, interpretability, and the organization of the latent space, as investigated in~\cite{}. Reconstruction ability captures how well the model is able to recover the sequence encoded in the latent space. The interpretability points out what properties are conserved in the latent space, for example by correlation off activity with Euclidean distance~\cite{}, cosine similarity~\cite{}, and by associating PCs with specific properties such as hydrophohicity~\cite{}. Finally, the organization of the latent space is understood as the ability to separate AMPs from non-AMPs, as well as locating regions of interests, for example with high density of highly active peptides. 

%In case of optimized generation, the models should be evaluated wrt to how well they tackled the Pareto optimization problem, by improving peptides towards two or more conflicting goals. 

%[Przewidywanie resistance na AMP]
\section*{Challenges and opportunities}
% \section*{Grand challenges and opportunities still lie ahead of AI method development for AMP design}
%Despite all the success of the extant AI approaches to AMP design,
Despite recent achievements in AI-driven AMP design, significant challenges still obscure discovery of clinically-relevant AMPs. %These include unavailability of clean, community-wide accepted training and benchmarking data, lack of discriminators and generation methods addressing relevant tasks, and lack of methods for ranking of the designed AMP sequences. 
%Each of these challenges present excellent opportunities for further development of the field,
These challenges offer prospects for field advancement, on the experimental, data collection and on the methodological side.
%What are the current main problems and what can be done to circumvent them? 

% Data noise
%As described previously, there is inherent noise in AMP data related to activity and toxicity measurements. The models need to handle the bias related to unit conversion as well as conflicting entries. The first and essential step in adressing that problem is reporting experimental conditions in detail, such as medium, pH, salt content and others. Second, unit conversion needs to be done carefully, taking into account the cell density as well as counter ion content. However, AI models could address the noise more explicitly. For example, Murakami et al.~\cite{murakami2023design} predict both the mean and the variance of the antimicrobial and hemolytic activities. In turn Huang et al.~\cite{huang2023identification} employ incremental learning framework, first training a MIC regression model, and later fine tuning it using additional MIC measurements taken in their lab. 
\paragraph{Construction of community-wide accepted benchmarking data}   %A contribution that would hugely benefit the entire field of AI design for AMPs would be benchmarking datasets. 
Antibacterial or hemolytic activity datasets should be carefully compiled from already existing, publicly available databases and include both positive AMP sequences that are active against given species of bacteria, as well as of negative (inactive) peptides. The construction of such datasets would require a series of pre-processing steps that would be acceptable to the entire research community, such as computing reasonable summary measures of activity, choosing a consensus activity threshold, and recording experimental conditions in detail, such as medium, pH, or salt content. Common use of clean, large benchmarking datasets, combined with open sharing of code, would foster reproducible method training and comparability.

\paragraph{Filling the current gaps of knowledge to enable novel method development}
Data on multiple  AMP properties, such as cytotoxicity, solubility, time to resistance, stability, or degradation %, not only do not come in clean benchmarking datasets, but more drastically, 
are either very rare or not present in databases at all. %, preventing AI model development. 
Activity measurements of clinical importance, such as MIC for %ESKAPE strains,
carbapenem-resistant \textit{Klebsiella pneumoniae}, \textit{Clostridium difficile}, \textit{Mycobacterium tuberculosis}, or New Delhi metallo-$\beta$-lactamase-producing strains, are most often not reported for the known AMPs. With MIC measurements against multiple strains, including those of clinical importance, a new type of %extremely useful 
generators could be developed that would be trained not only on AMP sequences but also on the bacterial genotypes in a multi-task manner, thereby boosting the size of the training set and bringing the functionality of the generator to a higher level. Although cytotoxicity data against different cell lines is already collected in DBAASP, it should be enlarged and  extended to alternative toxicity measures, such as nephrotoxicity and leucotoxicity. The acquisition of data on such alternatives would enable development 
%fuel the emergence 
of AI methods for accurately predicting or generating non-toxic peptides.
Currently, lipopeptides~\cite{cardoso2021molecular}, glycopeptides~\cite{cardoso2021molecular}, and peptoids~\cite{benjamin2022efficacy} or peptides modified using non-proteinogenic amino acids (NPAA) and retro amino acids are investigated as alternative to traditional peptides, but constitute only a minuscule fraction of database entries. 
Although first methods tailored to design peptides with NPAA~\cite{murakami2023design} 
%and N and C-terminal modifications~
were already proposed, 
%Interestingly, despite the clear influence of  on peptide structure and charge, only a limited number of methods have incorporated this information into their design\cite{sharma2023endl, losin2021exploring} 
further effort in this direction is needed to discover AMPs resistant to proteolytic degradation. 
%Existing methods consider solely standard amino acids, with an exception of MODAN~\cite{murakami2023design} 
%Development of AI methods accounting for NAA and N- and C- terminal modifications %are lacking and 
%would be promising step in finding AMPs resistant to proteolytic degradation. 
Moreover, the extreme scarcity of data %on clinically promising subcategories of AMPs other than antibacterial, such as anticancer AMPs, 
prevented development of reliable discriminators, and blocked development of anticancer AMP generators. 
To address the data scarcity issues and prompt AI-driven AMP discovery, 
% fuel emergence of novel AI approaches to AMP design, 
coordinated effort of large-scale initiatives would be required to systematically measure and report the data, akin to the Community for Open Antimicrobial Drug Discovery.  %~\cite{DOI: 10.1021/acsinfecdis.5b00044.}.   

% Data scarcity
%Another major limiting factor in AMP discovery is data scarcity. While the activity for reference strains and hemolytic measurements are fairly abundant, in case of truly dangerous strains, and cytotoxicity there is not enough data for model training purposes. 
\paragraph{Accounting for noise and scarcity of the data} In addition to
%the effort of collecting and systematically recording the data on the experimental and database side, 
data collection methods, equal emphasis should be laid on %methodological development to 
explicit modeling or correcting for the data noise and scarcity.
% developing and applying methodological approaches to explicit modeling or correcting for the noise inherent in the data and training from scarce samples.
While recent discriminators did attempt to tackle the data scarcity problem by transfer learning~\cite{lee2023ampbert, sharma2023artificial, salem2022ampdeep, jiang2023explainable, yu2021hmd, liu2023evolutionary},
%especially using large pre-trained language models[ref ref ref], 
%or by fuzzy~\cite{chharia2021novel}, zero-shot/one-shot learning~\cite{gull2020amp}, 
this problem has not been explicitly addressed by any of the AMP generation methods.  %have been made as well. While zero-shot and one-shot learning address the challenge of learning from limited labeled data or recognizing new classes, fuzzy learning explicitly deal with the uncertainty. 
% Importantly, in the presence of scarce and noisy both discriminators and generators should incorporate measures of model uncertainty to guide the users in utilizing their predictions or samples. 
Both discriminators and generators should incorporate model uncertainty measures to guide users in leveraging predictions or samples effectively.
%Another approach stems from microbial strain specific predictions, where additional genomic features are included to leverage inter-strain similarity. However, building separate models for each strain is suboptimal precisely due to the data scarcity. An alternative strategy could be multitask learning, or an recommender system, where multiple strains and multiple peptides and modeled at the same time. 

%[Benchmark, biased dataset - otworzyłam tym caly rozdzial]
%A thing influencing each and every AMP discovery method is the dataset construction. As every study follows their own experiment design it makes benchmarking and evaluation of the model a frustrating endeavor. In the light of this, sharing both code and datasets used for training should be a publishing prerequisite. However, a benchmarking dataset following the standards of ML community, such as CIFAR, needs to be constructed. 

\paragraph{Exploiting clustering of known AMPs} 
The antimicrobial peptide population exhibits notable similarity, as new antimicrobial peptides were frequently discovered by manipulating known AMPs, e.g. via  substitution of certain amino acids or cropping longer peptides to desired length. %While efforts were made previously to measure the similarity, e.g. by using the CD-HIT algorithm, they are usually targeted to remove that factor, e.g. to clean datasets from overly similar sequences, considering them as redundant. 
%Similarity of peptides is often evaluated using the CD-HIT algorithm~\cite{}, and datasets are usually cleaned from overly similar sequences, considering them as redundant. 
So far treated as redundant, the existence of matched pairs of original peptides and their synthetic analogues, opens up several modeling opportunities. First, models should leverage such similar pairs to learn the hard task of discriminating between peptides that are similar yet differ in activity, such as Brevinin-2 related peptide (MIC against \textit{E. coli} ATCC 25726 $=$ $20~\mu$M) and its analogue B2RP [K16L] which is five times weaker (MIC \textit{E. coli} ATCC 25726 $=$ $100~\mu$M), despite only one amino acid difference in sequence. In fact, filtering the similar peptides from the dataset overly simplifies the discrimination problem.  
%Moreover, different similarity cut-offs are applied, rendering model comparisons difficult. 
%An alternative approach based on expectation-maximization clustering was proposed by~\cite{pinacho2021alignment} to ensure a similar peptide distribution in both training and test data sets. Li et al.~\cite{li2022amplify} evaluate the performance of AMPlify model stratifying the test set based on sequence similarity. 
%In general, both activity/toxicity and sequence similarity needs to be included in definition of positive and negative examples, and model performance should be evaluated at various thresholds. 
% Finally, it is not straightforward how to choose the promising lead among a collection of peptides classified as AMP. 
Second, matched pairs should benefit the development of models for AMP comparison such as~\cite{losin2021exploring},  where two AMPs can be ordered by their property of interest. %Again, the inclusion of pairs that are highly similar in sequence yet different in property should largely benefit the training of such a comparator.
Finally, the clustering of peptides in the training data could be accounted for by generation methods, either by grouping peptide representations from the same cluster in the latent space, or attempting cluster-dependent, local organization of the latent.

\paragraph{Ranking of generated AMPs} %Finally, AMP discovery is simply expensive and slow, as experimental validation stands at its core. AI-assisted methods can be of help here in increasing the AMP success rate, especially by reducing the number of false positives. 
There is a dire need of approaches that rank the designed AMP sequences. Every generator can output thousands of sequences of novel AMP candidates, and only a limited number of those can be experimentally tested. Ideally, the candidates should be ranked by their chances of experimental validation. %: being highly active, having low toxicity, etc. 
This could be approached by applying existing discriminators to predict AMP properties for each candidate. However, many discriminative methods do not output a prediction score that could be used to rank the peptides, and many peptides rank similarly high according to the available scores. 
%The unconstrained and analogue generation methods often fail to provide such ranking as well. 
We have previously applied a conservative ensemble approach to filter and rank candidates based on predictions from multiple discriminators~\cite{szymczak2023discovering}. However, systematic, dedicated ensemble method development, both for ranking in terms of activity against different strains of bacteria, and toxicity, is still missing. %HA straightforward approach to the ranking would be creating an ensemble  resource collecting predictions from multiple discriminators. Thanks to each of them having an independent predictive bias, this could improve the prediction quality. 

%Apart from ranking, better filtration methods are needed as well.
\paragraph{AI-driven, faster and more specific molecular dynamics simulation methods} Another promising, but currently both time and computationally expensive direction in evaluating the quality of designed candidate AMP sequences is using fully atomistic simulations in the presence of bacterial or erythrocyte membrane. Here, AI could be leveraged to speed-up the molecular dynamics simulations, e.g. by predicting some simulation steps that effectively could be skipped. Similarly, AI-driven predictions of  peptide conformation in the presence of a cellular membrane could be used to develop more accurate discriminators and generators.  
%If MD simulations are to be used as ranking/filtering method, it is essential to simulate the conformation of candidate peptides in membrane environment. Additionally, it is crucial to consider both the starting conformation and the initial position of a peptide.  % Longer discussion of AlphaFold?

%Such a combination of efficient AI approaches and robotized experimentation %to peptide design, ranking, simulation-based evaluation, and and AI-equipped robotic experimentation
%could eventually provide the fastest way to design novel, potent AMPs. %Something about automated validation from small molecule design]. 
%What are important directions in AMP discovery?

\paragraph{Optimized generation methods of the future} Despite the recent  emergence of several  optimized AMP generation methods~\cite{jain2022biological, tucs2023quantum, hoffman2022optimizing}, this area has huge potential for future development. First, there exist no AI model that would be simultaneously trained in the task of both generation and optimization. Such a model would have the chance to exceed the optimisation capacity of current models, which %is largely limited. Indeed, no current generation model was 
so far were not able to design AMPs that would bear properties significantly better than known AMPs in the training set. %, such as AMPs with MIC $\textless$ 1 $\mu$g/mL. 
Only such AMPs have the potential to make their way to the clinic and replace current antibiotics. Moreover, since each peptide should be optimized in not one, but several properties, e.g. be simultaneously active and non-toxic, more effort should be invested in Pareto optimization approaches to AMP generation. Here, methods need to account for %recognizing 
the fact that increasing the activity and decreasing the toxicity of a peptide are in a way conflicting goals. %should lead to better success rate in AMP discovery. 
Finally, the optimized generation methods should account for a phenomenon that could be referred to as the {\textit{idealism-realism tradeoff}}. Namely, in the optimization process, the generated sequences should exceed the training samples in terms of their properties, but at the same time stay close to the training data, to remain biologically meaningful. The control of the tradeoff could be backed by additional estimators of peptide realism, e.g. 
by estimating relative ease of solid-state peptide synthesis using Milton Coupling Efficiency~\cite{ansari2023serverless}. 

\paragraph{AI-driven, accelerated, automated lab and design process}
Ultimately, the entire process of generating novel candidates, peptide synthesis, experimental validation in multiple parallel assays could be fully automated, and applied in an iterative, large-scale manner, potentially allowing for adaptive on-line improvement of both robots and AI methods. 
A rapid method of VAE-based generation and fast cell-free synthesis has been proposed to expedite peptide identification~\cite{pandi2022cell}, but other steps of the AI-driven AMP discovery pipeline should also be accelerated. 
%Jakieś conclusion?

%Additionally, it is crucial to develop methods that would predict activity not only against reference strains, but against strains (species?) of critical importance, such as ESKAPE strains, \textit{Clostridium difficile}, \textit{Mycobacterium tuberculosis}, or New Delhi metallo-$\beta$-lactamase-producing strains. Similarly, prediction of hemolytic activity needs to be extended to prediction of cytotoxicity, especially nephrotoxicity and leucotoxicity. For example, cytotoxicity measurements against various cell lines collected by DBAASP could be used for training such a model. 

%There is also a vast room for improvement in terms of predicting peptide stability and degradation. Currently, lipopeptides~\cite{}, glycopeptides~\cite{}, and peptoids~\cite{} are investigated as alternative to traditional peptides. Modifications using NAA and retro amino acids are researched as well. The emergence of AI methods addressing NAA and N- and C- terminal modifications is a promising step in finding AMPs resistant to proteolytic degradation. 

%An interesting research direction is prediction of anticancer peptides. As AMP were shown to selectively target cancer cells~\cite{}, there is a need for methods that would help discover peptides with anticancer properties but without the toxicity to the host system. 

\section*{Acknowledgments}

We would like to express our gratitude to Tomasz Grzegorzek, Adam Izdebski,  Michał Michalski, and Marcin Możejko for insightful discussions on the subject of AMP discovery.

\bibliography{bib.bib}{}
\bibliographystyle{ieeetr}

\end{document}